\documentclass{article}

\usepackage[pdftex]{graphicx, color}
\usepackage[colorlinks=true, linkcolor=black, filecolor=black, urlcolor=black, citecolor=black]{hyperref}
\usepackage{amsmath,graphicx,amssymb,amsfonts,bm,color,url,mathtools,cite}
\usepackage{multirow}
\usepackage{threeparttable}
\usepackage{flushend}
\allowdisplaybreaks[4]

\usepackage{authblk}
\usepackage{indentfirst}
\usepackage{geometry}
\allowdisplaybreaks[2]

\makeatletter
\long\def\@makecaption#1#2{%
  \normalsize
  \vskip\abovecaptionskip
  \sbox\@tempboxa{#1: #2}%
  \ifdim \wd\@tempboxa >\hsize
    #1: #2\par
  \else
    \global \@minipagefalse
    \hb@xt@\hsize{\hfil\box\@tempboxa\hfil}%
  \fi
  \vskip\belowcaptionskip}
\makeatother

\title{GE2E-AC: Generalized End-to-End Loss Training for Accent Classification}
\author[1]{Chihiro Watanabe\thanks{ch.watanabe@ntt.com}} 
\author[1]{Hirokazu Kameoka\thanks{This work was supported by JST CREST Grant Number JPMJCR19A3, Japan.}}
\affil[1]{{\normalsize NTT Communication Science Laboratories, 3-1, Morinosato Wakamiya, Atsugi-shi, Kanagawa Pref. 243-0198 Japan}}
\date{}

\begin{document}
\maketitle

\begin{abstract}
Accent classification or AC is a task to predict the accent type of an input utterance, and it can be used as a preliminary step toward accented speech recognition and accent conversion. Existing studies have often achieved such classification by training a neural network model to minimize the classification error of the predicted accent label, which can be obtained as a model output. Since we optimize the entire model only from the perspective of classification loss during training time in this approach, the model might learn to predict the accent type from irrelevant features, such as individual speaker identity, which are not informative during test time. To address this problem, we propose a GE2E-AC, in which we train a model to extract accent embedding or AE of an input utterance such that the AEs of the same accent class get closer, instead of directly minimizing the classification loss. We experimentally show the effectiveness of the proposed GE2E-AC, compared to the baseline model trained with the conventional cross-entropy-based loss.
\end{abstract}


\section{Introduction}
\label{sec:intro}

Accent is one of the major non-linguistic factors of speech. To alleviate the performance degradation in speech recognition caused by different accents (e.g., English, American, and Scottish), a potential solution is to first classify the accent type of an input utterance and then use an accent-specific speech recognition model that has been trained for the predicted accent \cite{Gupta82, Arslan96, Faria06}. Such a task to classify an accent type of a given utterance is called accent classification (AC), and various data sets have been publicly available for AC which contain the accent labels of the utterances as well as the speaker ones \cite{Demirsahin20, Sanabria23, Wang24}.

\begin{figure*}[t]
\begin{center}
\includegraphics[height=4.5cm]{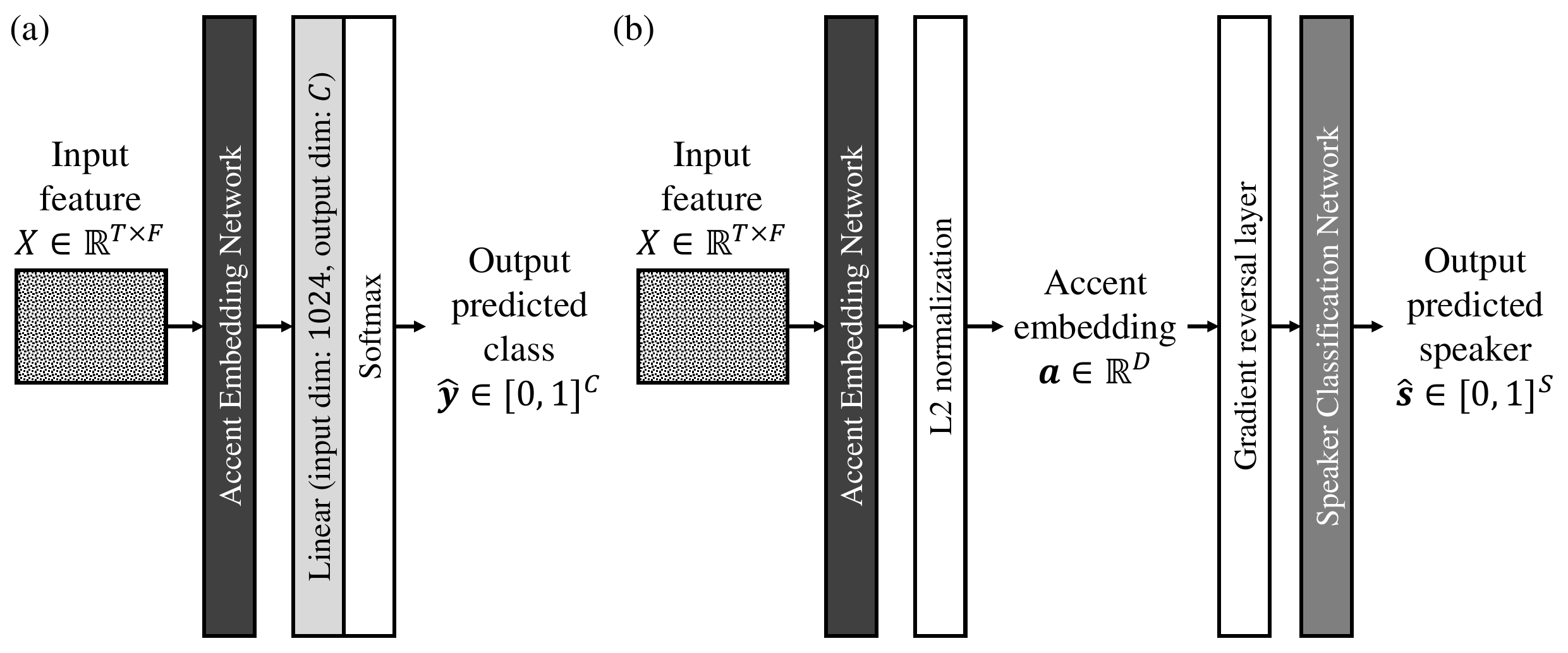} 
\includegraphics[height=4.5cm]{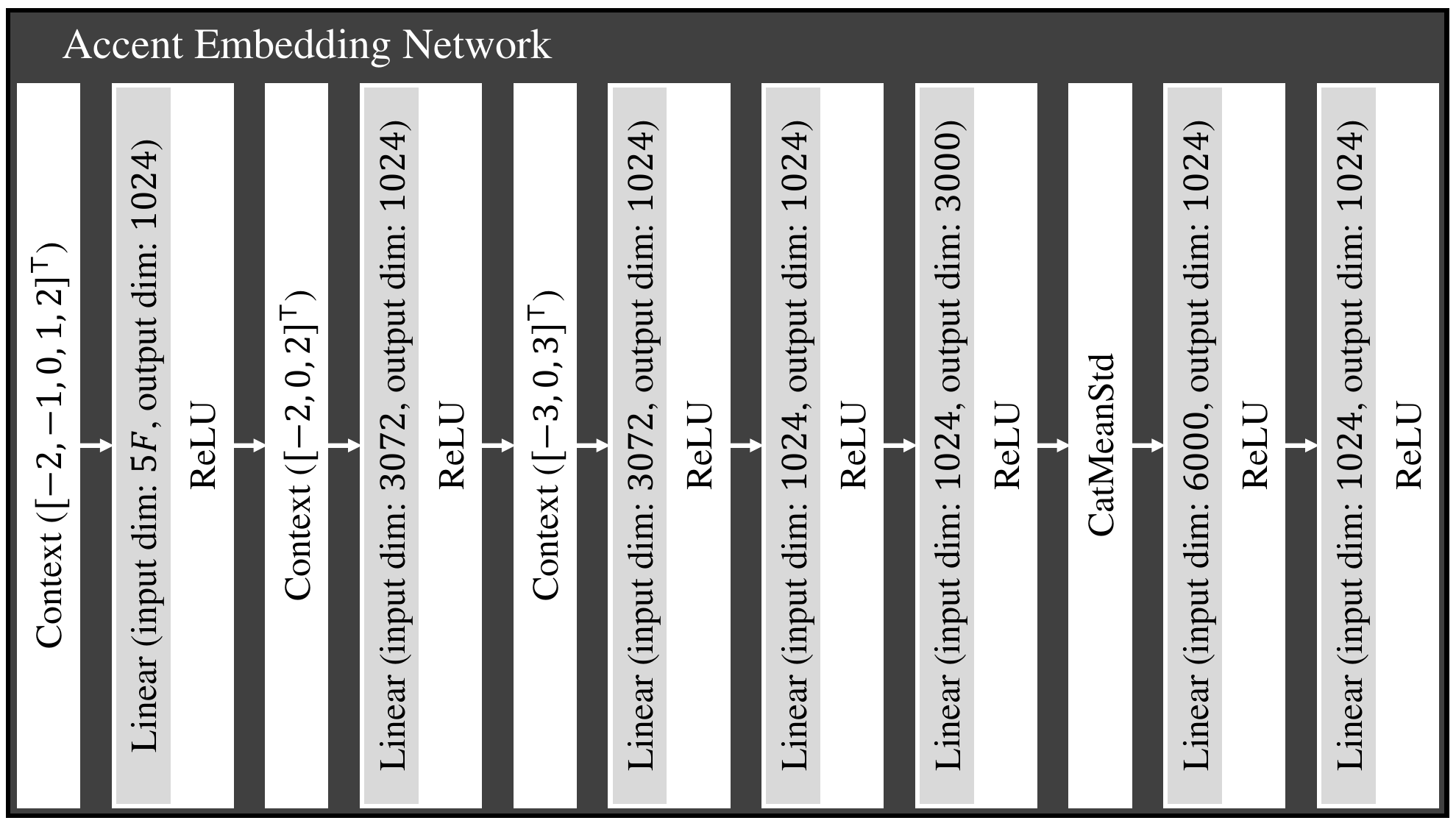}
\includegraphics[height=4.5cm]{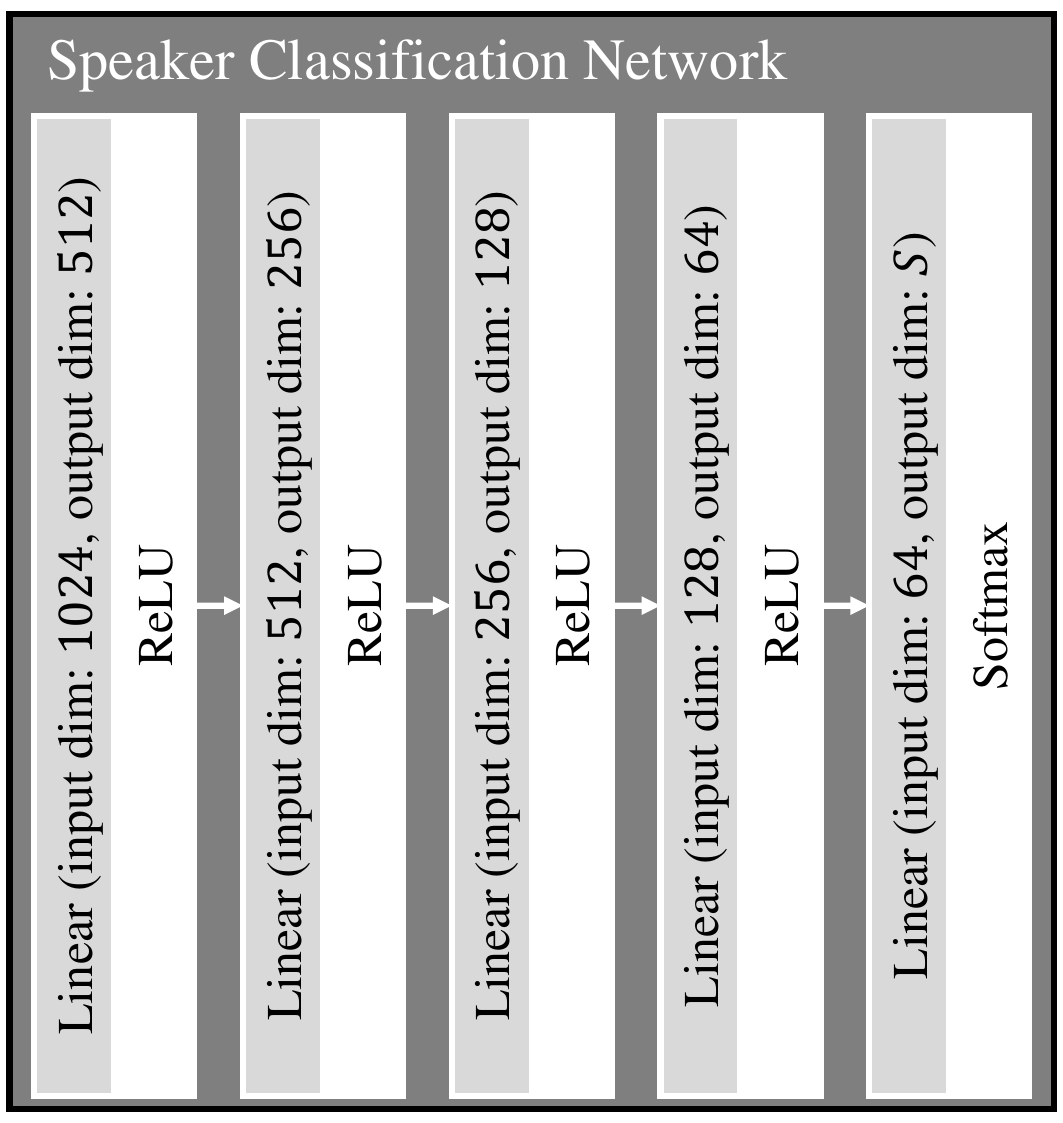}
\end{center}
\caption{The network architectures of (a) the baseline CE-based and (b) the proposed GE2E-based accent classification. For arbitrary vector $\bm{\phi} \in \mathbb{R}^K$, $\mathrm{Context}(\bm{\phi})$ indicates a function $f_{\bm{\phi}}: \mathbb{R}^{T \times F'} \mapsto \mathbb{R}^{T \times KF'}$, which maps matrix $X \in \mathbb{R}^{T \times F'}$ to $f_{\bm{\phi}}(X) = \left[ M^{(1)} X, \cdots, M^{(K)} X \right]$, where $M^{(k)} \in \mathbb{R}^{T \times T}$ is a matrix each of whose entries $M^{(k)}_{ij}$ is one if $j = i + \phi_k$ and zero otherwise for all $k \in \{ 1, \dots, K \}$. $\mathrm{CatMeanStd}$ is a function $f: \mathbb{R}^{T \times F'} \mapsto \mathbb{R}^{2F'}$, which maps matrix $X \in \mathbb{R}^{T \times F'}$ to $\left[\bm{\mu}^\mathsf{T}, \bm{\sigma}^\mathsf{T}\right]^\mathsf{T}$, where $\bm{\mu} \in \mathbb{R}^{F'}$ and $\bm{\sigma} \in \mathbb{R}^{F'}$ are the vectors each of whose entries $\mu_k$ and $\sigma_k$ are the mean and the standard deviation of vector $\left[ X_{1k}, \dots, X_{Tk} \right]^\mathsf{T}$, respectively.}
\label{fig:network}
\end{figure*}

Existing AC models are typically trained to minimize a loss function (e.g., cross-entropy or CE loss) between the target and predicted accent labels \cite{Jain18, Lesnichaia22, Nechaev24}. We will call this approach \textbf{CE-AC} (Fig. \ref{fig:network} (a)). While this approach directly optimizes the model's classification accuracy during training, it may lead to predictions that capture irrelevant information (e.g., speaker identity) from the input utterance to reduce the training loss, ultimately degrading test performance. This has also been pointed out in \cite{Shi21}. 
Moreover, CE-based approach is based on the implicit assumption that the set of potential accent types of input utterances is fixed. To utilize an additional data set with the accent types that have not been in the previous data set, a different model with a new output size must be retrained from scratch.

To address these limitations, we propose adapting the concept of generalized end-to-end (GE2E) loss training \cite{Wan18}, which has proven highly effective in speaker verification tasks, for use in the AC task. Specifically, instead of training a model to directly minimize the error in the predicted accent class, we train it to extract accent embeddings (AEs), i.e., feature vectors representing accent information, from input utterances so that the AEs of the same accent class get closer, while those of different accent classes move farther apart. We call this approach \textbf{GE2E-AC} (Fig. \ref{fig:network} (b)).

This approach has the advantage that even when a test sample with an unknown accent comes in, it can determine which training sample is most similar. Additionally, it is easy to retrain without needing to modify the model structure. 
Our experiments show that this method is more effective for the AC task compared to the conventional CE-based baseline method. In the experiment, we also compared the basic GE2E-AC with \textbf{GE2E-AC-A}, which incorporates the additional adversarial speaker classification loss to further eliminate the speaker identity information from extracted AEs. 
Furthermore, since the optimal input for the AC task is not obvious, we also evaluated the performance of using phoneme-related bottle-neck features (BNFs) \cite{Sun16} extracted from an encoder-decoder type automatic speech recognition (ASR) system and Hidden-Unit BERT (HuBERT) \cite{Hsu21} features as inputs.

\section{Baseline method: CE-based accent classification}
\label{sec:baseline}

Before we explain the proposed GE2E-based AC, we first describe a baseline CE-AC. In this approach, we use accent classifier $f_{\mathrm{CE}}: \mathbb{R}^{T \times F} \mapsto \mathbb{R}^C$ which maps input feature $X \in \mathbb{R}^{T \times F}$ (e.g., BNF \cite{Sun16} and HuBERT \cite{Hsu21}) of an utterance to vector $\hat{\bm{y}} = f_{\mathrm{CE}}(X) \in [0, 1]^C$, where $T$, $F$, and $C$ indicate the numbers of time and frequency bins of an input feature and accent types, respectively. Throughout this paper, we assume that classifier $f_{\mathrm{CE}}$ is modeled as a neural network. By assuming that output vector $\hat{\bm{y}}$ represents the probabilities of categorical distribution on the accent types, we train the entire network $f_{\mathrm{CE}}$ by minimizing the CE loss $\mathcal{L}_{\mathrm{CE}}$ between the model output $\hat{\bm{y}}$ and the target accent label in one-hot representation $\bm{y} \in \left\{ \bm{y}' \in \{0, 1\}^C | \sum_{j=1}^C y'_j = 1 \right\}$. For a mini-batch with size $B$, the CE loss is given by
\begin{align}
\label{eq:loss_ce}
\mathcal{L}_{\mathrm{CE}} = -\frac{1}{B} \sum_{l=1}^B \sum_{j=1}^C y_{lj} \log \hat{y}_{lj},
\end{align}
where $\bm{y}_l = (y_{lj})_{1 \leq j \leq C}$ and $\hat{\bm{y}}_l = (\hat{y}_{lj})_{1 \leq j \leq C}$ are the target and predicted accent labels of the $l$th sample in a mini-batch, respectively.

In the experiment in Section \ref{sec:experiment}, we used a classifier based on the architecture in \cite{Snyder18}, as shown in Fig. \ref{fig:network} (a). This architecture is also used as a sub-model in \cite{Huang21}, which achieved the highest accuracy in track 1 (i.e., English accent recognition task) of the accented English speech recognition challenge 2020 (AESRC2020) \cite{Shi21}.


\section{Proposed method: GE2E-based accent classification}
\label{sec:proposed}

\subsection{GE2E-AC}

During training, the baseline method described in Section \ref{sec:baseline} focuses on directly minimizing the error in the predicted accent label. However, this training approach may cause the model to rely on irrelevant information, such as speaker identity, to reduce training loss, leading to poor test performance. To prevent this, we propose using a metric learning approach that trains the model to extract AEs from input utterances, ensuring that the AEs of utterances with the same accent class are closer together while those of different accent classes are farther apart. This approach also offers significant advantages in potential applications beyond improving the model's generalization ability. Specifically, the ability to extract AEs allows it to be used for tasks like zero-shot accent conversion, where input speech is converted to match the accent of a reference speech sample, even if the reference accent is unknown.

GE2E loss training is a type of metric learning approach originally used for speaker verification \cite{Wan18}. It has proven highly effective in this task by extracting speaker embeddings from input utterances so that embeddings from the same speaker are closer together, while those from different speakers are farther apart. In our proposed method, we adapt this approach for the AC task, treating accent classes similar to how speaker identities are handled in speaker verification. In the proposed GE2E-AC, by pulling the AEs from the utterances of the same accent type closer together and pushing those from differently accented utterances away, the model would learn to extract an abstract representation of accent, thereby avoiding overfitting to individual speakers. 
Although there are other possible methods of deep metric learning such as Siamese network \cite{Siddhant17}, we employ GE2E loss here, since its effectiveness has already been shown in another speech processing task \cite{Wan18}.

The actual flow of GE2E-AC is as follows (Fig. \ref{fig:network} (b)). We first extract a $D$-dimensional AE $\bm{a} = f_{\mathrm{AE}}(X) \in \mathbb{R}^D$ from input feature $X$ by using an accent embedding network $f_{\mathrm{AE}}: \mathbb{R}^{T \times F} \mapsto \mathbb{R}^D$. During training, we define a mini-batch of input features as a set of $M$ utterances for each of $C$ accent types. 
Let $\bm{a}_{ji} \in \mathbb{R}^D$ be the AE of the $i$th utterance of the $j$th accent for $i = 1, \dots, M$, and $j = 1, \dots, C$. As shown in Fig. \ref{fig:network} (b), we define an AE such that it has a unit L2 norm (i.e., $\|\bm{a}_{ji}\|_2 = 1$). Based on these AEs, we can compute two types of centroids for each accent class $j$, which are given by
\begin{align}
\bm{c}_j = \frac{1}{M} \sum_{i=1}^M \bm{a}_{ji}, \ \ \ 
\bm{c}^{(-i)}_j = \frac{1}{M-1} \sum_{\substack{m=1\\m \neq i}}^M \bm{a}_{ji}.
\end{align}
We also define the L2 normalization of the centroids by
\begin{align}
\bar{\bm{c}}_j = \frac{\bm{c}_j}{\|\bm{c}_j\|_2}, \ \ \ 
\bar{\bm{c}}^{(-i)}_j = \frac{\bm{c}^{(-i)}_j}{\|\bm{c}^{(-i)}_j\|_2}.
\end{align}
By using these L2 normalized centroids, we define $MC \times C$ similarity matrix $S = (S_{lk})_{1 \leq l \leq MC, 1 \leq k \leq C}$, whose $(l, k)$th entry is given by
\begin{align}
\label{eq:sim_matrix}
S_{lk} = \begin{cases}
w \bm{a}_{ji} \cdot \bar{\bm{c}}^{(-i)}_k + b & j=k, \\
w \bm{a}_{ji} \cdot \bar{\bm{c}}_k + b & \mathrm{otherwise},
\end{cases}
\end{align}
where $w$ and $b$ are learnable weight and bias parameters for the similarity metric and $l = (j-1)M + i$ for $i = 1, \dots, M$ and $j = 1, \dots, C$. As in \cite{Wan18}, we define each entry $S_{lk}$ differently in the cases of $j=k$ and $j \neq k$ for training stability.

Based on similarity matrix $S$ in \eqref{eq:sim_matrix}, we train the entire network to minimize the GE2E loss, which is given by
\begin{align}
\label{eq:loss_ge2e}
\mathcal{L}_{\mathrm{GE2E}} = -\frac{1}{MC} \sum_{l=1}^{MC} \sum_{k=1}^C y_{lk} \log \hat{y}_{lk}, 
\end{align}
where $\hat{y}_{lk} = \exp(S_{lk}) / \{\sum_{k'=1}^C \exp(S_{lk'})\}$. In regard to the final formulation, the GE2E loss in \eqref{eq:loss_ge2e} looks similar to the CE loss in \eqref{eq:loss_ce}. However, unlike in the CE loss, the predicted accent labels $\{\hat{y}_{lk}\}$ in the GE2E loss are computed based on the similarity $\{S_{lk}\}$ between the AEs and their centroids. To minimize the GE2E loss, the accent embedding network should learn to pull the AEs closer to the centroid of its accent type and to push them away from the other centroids. This facilitates the AEs of the same accent type getting closer together, even if they are uttered by mutually different speakers, which would improve the generalization performance in AC.

In the experiment in Section \ref{sec:experiment}, to evaluate the effect of introducing the GE2E loss instead of the CE one, we set the shallower part of the model (i.e., ``Accent Embedding Network'' in Fig. \ref{fig:network}) to be the same as that of CE-AC. In the standard GE2E-AC, we used only the shallower part than the gradient reverse layer.

\subsection{GE2E-AC-A}

Although introduction of GE2E loss alone would improve the performance of AC during test time, AEs of GE2E-AC might still contain some information related to speaker identity, since there are no samples of different accents from the same speaker. To further eliminate speaker-related information from AEs and improve the generalization performance, we also propose GE2E-AC-A, in which we incorporate an adversarial speaker classification or SC loss $\mathcal{L}_{\mathrm{SC}}$ into the training criterion. We consider such a loss to avoid mixing speaker information in AEs. It must be noted that a similar technique has also been used in \cite{Zhou24}, although our main contribution lies in introduction of GE2E loss for AEs, rather than in that of SC loss. 

Specifically, the SC loss can be computed by using another network $f_{\mathrm{SC}}: \mathbb{R}^D \mapsto [0, 1]^S$ (i.e., ``Speaker Classification Network'' in Fig. \ref{fig:network}), which predicts a speaker index from an AE. Let $\bm{s}\in \left\{ \bm{s}' \in \{0, 1\}^S | \sum_{j=1}^S s'_j = 1 \right\}$ be the target speaker label in one-hot representation and let $\hat{\bm{s}} = f_{\mathrm{SC}}(\bm{a}) \in [0, 1]^S$ be the predicted one. The SC loss is given by
\begin{align}
\label{eq:loss_sc}
\mathcal{L}_{\mathrm{SC}} = -\frac{1}{B} \sum_{l=1}^B \sum_{j=1}^S s_{lj} \log \hat{s}_{lj},
\end{align}
where $\bm{s}_l = (s_{lj})_{1 \leq j \leq S}$ and $\hat{\bm{s}}_l = (\hat{s}_{lj})_{1 \leq j \leq S}$ are the target and predicted speaker labels of the $l$th sample in a mini-batch, respectively. 

The training strategy of GE2E-AC-A is two-fold. The speaker classification network should be trained to minimize SC loss $\mathcal{L}_{\mathrm{SC}}$, while the accent embedding network should be trained to maximize it. We can implement such an adversarial learning by adding a gradient reversal layer (GRL) \cite{Ganin16} before the speaker classification network. The GRL is an identity function in forward propagation, while it multiplies the gradient from the deeper layers by $-1$ in backpropagation. By introducing GRL, we can formulate the overall training criterion of GE2E-AC-A to be minimized by $\mathcal{L}_{\mathrm{GE2E}} + \lambda_{\mathrm{SC}} \mathcal{L}_{\mathrm{SC}}$, where $\lambda_{\mathrm{SC}}$ is a hyperparameter. As GE2E-AC, we used a network architecture in Fig. \ref{fig:network} (b) in the experiment in Section \ref{sec:experiment}.


\section{Experiment}
\label{sec:experiment}

To confirm the effectiveness of the proposed GE2E-AC, we conducted an experiment using CSTR VCTK Corpus \cite{Veaux17}. Among this data set, we only used the samples of accent types of ``English,'' ``American,'' ``Scottish,'' ``Irish,'' and ``Canadian,'' which consist of $33$, $22$, $19$, $9$, and $8$ speakers, respectively. For each accent type, we randomly chose five test speakers and used the other speakers for training. 
We trimmed leading and trailing silence of each sample and resampled it to $16$ KHz. 

As input features, we adopted BNF \cite{Sun16} and HuBERT \cite{Hsu21}. We extracted these features from WAV samples by using the codes provided by \cite{Liu21, huberthf} and trained each model of CE-AC (baseline), GE2E-AC, and GE2E-AC-A with them.

The hyperparameters for training the models were as follows. For all the models, we used Adam \cite{Kingma15} with the learning rate of $1.0 \times 10^{-5}$ for optimization and set the minibatch size to $16$, the number of epochs to $100$, $\lambda_{\mathrm{SC}}$ to $1.0 \times 10^{-5}$, and the number $M$ of utterances per accent type to $10$. 
It must be noted that, since a mini-batch of GE2E-AC and GE2E-AC-A consists of the same number $M$ of utterances for all the accents with generally different sample sizes $\{n_1, \dots, n_C\}$, we could not determine the meaning of one epoch in the usual sense. Therefore, for these methods, we defined the nominal ``sample size'' for a sample which includes one input feature for each accent as the maximum sample size $\max_{j=1, \dots, C} n_j$ in the experiment. 
We clipped the max L2 norm of the gradients of the model parameters to $1.0$ for CE-AC and $3.0$ for GE2E-AC and GE2E-AC-A. Finally, we computed confusion matrices by using the trained models. The $(i, j)$th entry of a confusion matrix indicates the number of training or test samples whose target accent index is $i$ and whose predicted counterpart is $j$. 
Since outputs of the GE2E-based models are accent embeddings $\bm{a}$, not predicted accent classes $\hat{\bm{y}}$, we need to determine $\hat{\bm{y}}$ from $\bm{a}$ to compute a confusion matrix. Based on the accent embeddings of all the training samples, we first computed the centroid (i.e., mean of embeddings) of each accent class. 
We defined the predicted accent class of a given sample with accent embedding $\bm{a}'$ as the class whose centroid had the highest cosine similarity with $\bm{a}'$.

\begin{figure*}[p]
\begin{center}
\includegraphics[width=\linewidth]{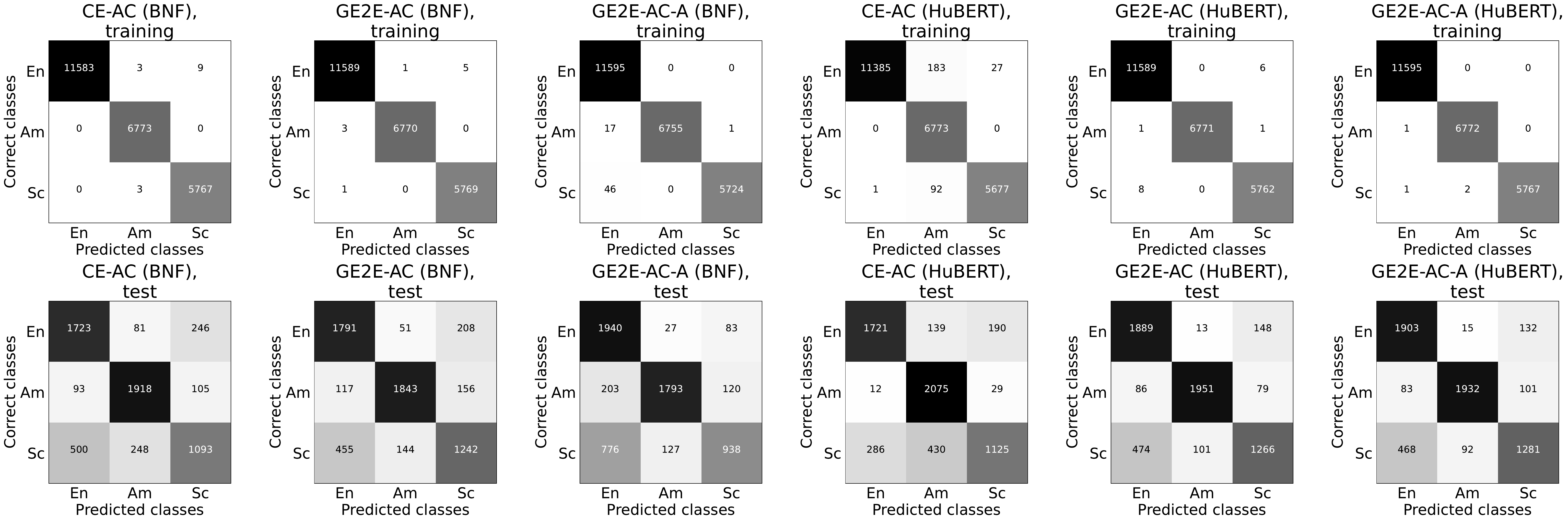}
\end{center}
\caption{Confusion matrices with \textbf{three} accent types. The value plotted in each entry indicates the number of corresponding samples. ``En,'' ``Am,'' and ``Sc'' stand for English, American, and Scottish, respectively.}
\label{fig:conf3}
\begin{center}
\includegraphics[width=\linewidth]{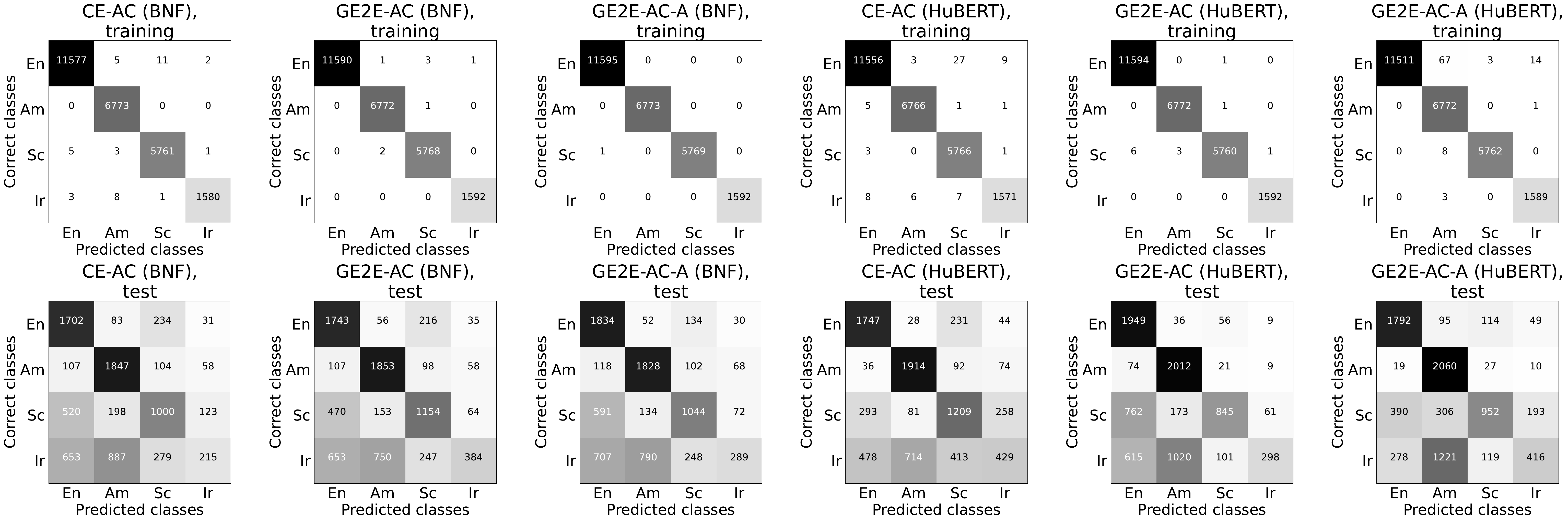}
\end{center}
\caption{Confusion matrices with \textbf{four} accent types. ``Ir'' stands for Irish.}
\label{fig:conf4}
\begin{center}
\includegraphics[width=\linewidth]{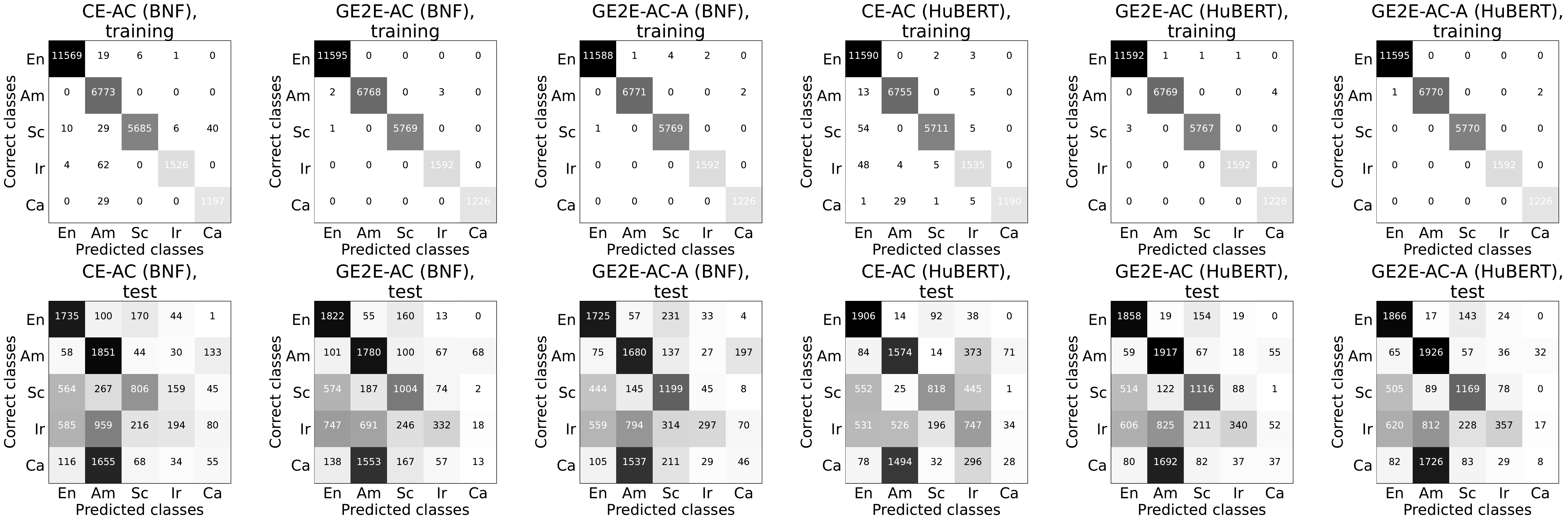}
\end{center}
\caption{Confusion matrices with \textbf{five} accent types. ``Ca'' stands for Canadian.}
\label{fig:conf5}
\end{figure*}

Tables \ref{tab:acc3}, \ref{tab:acc4}, and \ref{tab:acc5} show the classification accuracy of each method in the cases of three, four, and five accent types, respectively. Aside from the case of using HuBERT for classifying four accent types, the proposed GE2E-AC or GE2E-AC-A achieved better test performance than that of the CE-based baseline method. In some cases (e.g., using HuBERT for classifying four accent types), the adversarial speaker classification loss in combination with the GE2E loss was effective, while in other cases it was not. In regard to the input feature type, HuBERT yielded better test performance than BNF did in most cases. As more detailed results, Figs. \ref{fig:conf3}, \ref{fig:conf4}, and \ref{fig:conf5} show the confusion matrices in the cases of three, four, and five accent types, respectively. Although there were slight differences in the values of the confusion matrices, they showed similar classification tendency to each other among all the settings. For instance, when the number of accent types was five, it was hard for all the models to correctly classify the Canadian speakers, and most of the utterances by Canadian speakers were classified as ``American.'' This would be probably due to the bias in sample size, and to mitigate such a problem would be an important direction of a future study.

\begin{table}[t]
\centering
\caption{Accuracy of accent recognition for each setting with \textbf{three} accent types.}
\label{tab:acc3}
\begin{tabular}{lcc}
Method (Feature type) & Training accuracy & Test accuracy \\ \hline
CE-AC (BNF)       & $99.9\%$ & $78.8\%$ \\ \hline
GE2E-AC (BNF)        & $100.0\%$ & $\mathbf{81.2\%}$ \\ \hline
GE2E-AC-A (BNF)     & $99.7\%$ & $77.8\%$ \\ \hline
CE-AC (HuBERT)    & $98.7\%$ & $81.9\%$ \\ \hline
GE2E-AC (HuBERT)     & $99.9\%$ & $85.0\%$ \\ \hline
GE2E-AC-A (HuBERT)  & $100.0\%$ & $\mathbf{85.2\%}$ \\ \hline
\end{tabular}
\caption{Accuracy of accent recognition for each setting with \textbf{four} accent types.}
\label{tab:acc4}
\begin{tabular}{lcc}
Method (Feature type) & Training accuracy & Test accuracy \\ \hline
CE-AC (BNF)       & $99.8\%$ & $59.2\%$ \\ \hline
GE2E-AC (BNF)        & $100.0\%$ & $\mathbf{63.8\%}$ \\ \hline
GE2E-AC-A (BNF)     & $100.0\%$ & $62.1\%$ \\ \hline
CE-AC (HuBERT)    & $99.7\%$ & $\mathbf{65.9\%}$ \\ \hline
GE2E-AC (HuBERT)     & $100.0\%$ & $63.5\%$ \\ \hline
GE2E-AC-A (HuBERT)  & $99.6\%$ & $64.9\%$ \\ \hline
\end{tabular}
\caption{Accuracy of accent recognition for each setting with \textbf{five} accent types.}
\label{tab:acc5}
\begin{tabular}{lcc}
Method (Feature type) & Training accuracy & Test accuracy \\ \hline
CE-AC (BNF)       & $99.2\%$ & $46.6\%$ \\ \hline
GE2E-AC (BNF)        & $100.0\%$ & $\mathbf{49.7\%}$ \\ \hline
GE2E-AC-A (BNF)     & $100.0\%$ & $49.6\%$ \\ \hline
CE-AC (HuBERT)    & $99.4\%$ & $50.9\%$ \\ \hline
GE2E-AC (HuBERT)     & $100.0\%$ & $52.8\%$ \\ \hline
GE2E-AC-A (HuBERT)  & $100.0\%$ & $\mathbf{53.4\%}$ \\ \hline
\end{tabular}
\end{table}

\section{Conclusion}
\label{sec:conclusion}

In this study, we proposed a new training method for AC based on the similarity between a pair of accent embeddings of the utterances so that the model can avoid overfitting to specific speakers. The effectiveness of the proposed GE2E-AC has been shown experimentally, compared to the existing CE-based method in which we directly minimize the classification loss for individual utterances. A possible future direction of this study would be to utilize the proposed method as a preliminary step toward accent conversion, where we aim to convert the accent of an input utterance to some different one. By extracting an AE from a target sample with a trained GE2E-AC model and using it for test-time accent conversion, it would be possible to convert accents in zero-shot manner.


\clearpage
\bibliographystyle{abbrv}
\bibliography{template}

\end{document}